\documentclass[11pt,reqno,a4,onecolumn,twoside,final,spanish]{article}

\usepackage{amsthm,amsmath,amssymb,latexsym,amsfonts,amscd} 
\usepackage{latexsym}
\usepackage{url}
\usepackage[T1]{fontenc}
\usepackage{ae}
\usepackage{aecompl}
\usepackage[small,sc,hang,bf]{caption} 
\usepackage{graphicx,color,ae,fancyvrb}
\usepackage{pst-all}
\usepackage{subfig}
\usepackage{float} 
\usepackage{verbatim}
\usepackage{enumerate}
\usepackage{array}
\usepackage{longtable}
\usepackage{multicol}
\usepackage{cancel}
\newcommand*\chancery{\fontfamily{pzc}\selectfont}

\newlength\mytemplen
\newsavebox\mytempbox

\makeatletter
\newcommand\mybluebox{%
    \@ifnextchar[
       {\@mybluebox}%
       {\@mybluebox[0pt]}}

\def\@mybluebox[#1]{%
    \@ifnextchar[
       {\@@mybluebox[#1]}%
       {\@@mybluebox[#1][0pt]}}

\def\@@mybluebox[#1][#2]#3{
    \sbox\mytempbox{#3}%
    \mytemplen\ht\mytempbox
    \advance\mytemplen #1\relax
    \ht\mytempbox\mytemplen
    \mytemplen\dp\mytempbox
    \advance\mytemplen #2\relax
    \dp\mytempbox\mytemplen
    \colorbox{myblue}{\hspace{1em}\usebox{\mytempbox}\hspace{1em}}}

\makeatother

\newcommand\D{\delta}

\newcommand\F{\phi}

\newcommand\R{\rho}

\newcommand\mh{\mathcal{H}}

\newcommand\nc{\nabla^2}

\usepackage{natbib}

\usepackage{hyperref}

\usepackage{geometry}
\usepackage{calc}
	
\makeatletter
\renewcommand{\section}{\@startsection{section}{1}{0pt}{-3ex plus -1ex minus 0ex}{2ex plus 0ex}{\bf}}
\makeatother

\makeatletter
\renewcommand{\subsection}{\@startsection{subsection}{1}{0pt}{-2ex plus -1ex minus 0ex}{2ex plus 0ex}{\bf}}
\makeatother

\theoremstyle{definition}

\theoremstyle{remark}

\begin{document}

\renewcommand{\tablename}{Tabla}
\renewcommand{\figure}{Figura}
\noindent

\begin{flushleft}
\textsl {\chancery  Memorias de la Primera Escuela de Astroestad\'istica: M\'etodos Bayesianos en Cosmolog\'ia}\\
\vspace{-0.1cm}{\chancery  9 al 13 Junio de 2014.  Bogot\'a D.C., Colombia }\\
\textsl {\scriptsize Editor: H\'ector J. Hort\'ua}\\
\href{https://www.dropbox.com/sh/nh0nbydi0lp81ha/AACJNr09cXSEFGPeFK4M3v9Pa}{\tiny {\blue Material suplementario}}
\end{flushleft}



\thispagestyle{plain}\def\@roman#1{\romannumeral #1}



\begin{center}\Large\bfseries Implicaciones Cosmol\'ogicas del Colapso Gravitacional en Teor\'ias de  $f(R)$ \end{center}
\begin{center}\normalsize\bfseries  Cosmological implications of gravitational collapse in $F(R)$ theories \end{center}

\begin{center}
\small
\textsc{C\'esar  Peralta \footnotemark[1]}
\textsc{Leonardo Casta\~neda \footnotemark[2]}
\footnotetext[1]{Universidad Nacional de Colombia. E-mail: \url{cdperaltag@unal.edu.co}}
\footnotetext[2]{Universidad Nacional de Colombia. E-mail: \url{lcastanedac@unal.edu.co}}


\end{center}

\noindent\\[1mm]
{\small
\centerline{\bfseries Resumen}\\
Haremos una comparaci\'on entre la din\'amica del colapso gravitacional esf\'erico para un universo de Friedmann-Lema\^{i}tre-Robertson-Walker (FLRW) de primer orden en el contexto de relatividad general, con el resultado correspondiente obtenido para el colapso gravitacional en el marco de la teor\'ia de gravedad modificadad $f(R)$.
Este trabajo tiene como objetivo la creaci\'on de un modelo anal\'itico que explique  las fuentes de las estructuras a grandes escalas en el universo presentando una aproximaci\'on del modelo de colapso gravitacional esf\'erico dentro del marco de las teor\'ias de gravedad modificada $f(R)$.\\

{\footnotesize
\textbf{Palabras clave: Colapso Gravitacional, Relatividad General, Teorias de  $f(R)$}\\
\noindent\\[1mm]
{\small
\centerline{\bfseries Abstract}\\

We will make a comparison between the dynamics of spherical gravitational co\-llap\-se for a perturbed FLRW universe to first order in the context of general relativity, with the corresponding results obtained for the gravitational collapse under theories of modified gravity f(R).
This work is aimed at obtaining an analytical model of explanation a source of large scale structures in the universe presenting an approximation of model spherical gra\-vi\-ta\-tio\-nal collapse under theories of modified gravity f(R). \\

{\footnotesize
}

\newpage
\section{Introducci\'on}

La teor\'ia de la relatividad general de Einstein se constituye como el paradigma cient\'ifico m\'as importante en el \'ambito de la descripci\'on gravitacional por sus aportes conceptuales, soportes experimentales de alta precisi\'on en los diferentes test de eva\-lua\-ci\'on que la teor\'ia misma ha promovido y, recientemente, resultados en el \'ambito de la simulaci\'on computacional que nos presenta un panorama virtual casi indistinguible al ojo humano de las observaciones m\'as recientes del universo. Al mismo tiempo es, junto con la mec\'anica cu\'antica, la teor\'ia m\'as exacta j\'amas antes estudiada en toda la historia de la ciencia. De igual modo, presenta una descripci\'on coherente del espacio, el tiempo, gravedad y materia a un nivel macrosc\'opico. En el \'ambito de la cosmolog\'ia, la teor\'ia de la relatividad general le da un nuevo soporte, dado que las ideas fundamentales sobre las percepciones del universo se van abriendo a nuevas posibilidades de exploraci\'on, medici\'on y mo\-de\-la\-mien\-to. Desde el inicio de esta teor\'ia, Eins\-tein se pregunt\'o si \'esta era la teor\'ia definitiva capaz de describir las interacciones gravitacionales. Comenz\'o as\'i la carrera de verificar y probar la relatividad general. En las \'ultimas tres d\'ecadas se empezaron a mostrar fallas fundamentales de la teor\'ia   en lo que respecta a la descripci\'on de los problemas cosmol\'ogicos (Problema de planitud, monopolos magn\'eticos, paredes de dominio, horizonte) mediante el modelo de Big Bang, para lo cual se deriv\'o un modelo que resuelve estos problemas en un escenario c\'osmico llamado inflaci\'on  y la incapacidad de dar una descripci\'on cu\'antica de la gravedad, \cite{AHguth}. Estas razones han dado paso a varias propuestas te\'oricas para conseguir modelos que puedan reproducir las interacciones gravitacionales en estas escalas. Una de las m\'as \'utiles presentadas son las teor\'ias de gravedad extendidas, \cite{scapozziellofranca}, las cuales est\'an basadas en las correciones y ampliaciones a la teor\'ia de Einstein. Dentro de estas teor\'ias alternativas, se encuentran las que permiten que el Lagrangiano dependa de una familia de diferentes funciones del escalar de Ricci, conocidas como teor\'ias de gravedad modificada $f(R)$.

Se encuentran en la literatura varias motivaciones para modificar la relatividad general. Entre ellas se  incorporar completamente el principio de Mach en la teor\'ia. Einstein s\'olo pudo introducir una de las ideas de Mach y a pesar de ello, esta idea admite una soluci\'on anti-Machaiana, el universo de G\"{o}del, \cite{godel}. De acuerdo con el principio de Mach, el marco de referencia inercial local est\'a determinado por el movimiento promedio de los objetos astron\'omicos distantes. Esta caracter\'istica implica que el acople gravitacional en un punto del espacio-tiempo no es absoluta, sino que est\'a determinada por la materia circundante. Por tanto,  se convierte en una funci\'on de la ubicaci\'on en el espacio-tiempo -un campo escalar- es decir, la constante gra\-vi\-ta\-cio\-nal cambia de acuerdo a la \'epoca en la que se encuentre el universo. La teor\'ia de Brans-Dicke fue la primera en plasmar completamente una alternativa a la relatividad general de Einstein dentro del marco de teor\'ias de gravedad alternativa. Consigui\'o la variaci\'on del acople gravitacional acoplando un campo escalar no m\'inimalmente a la geometr\'ia consiguiendo una implementaci\'on m\'as adecuada del principio de Mach que en el caso de relatividad general \cite{bransdicke}.

Por otro lado, las teor\'ias de gravedad extendidas exhiben naturalmente un comportamiento inflacionario capaz de sobrellevar las fallas del modelo de Big Bang basado en relatividad general \cite{starobinsky}. Se ha demostrado que, por medio de transformaciones conformes, t\'erminos de alto orden en la curvatura y t\'erminos acoplados no m\'inimalmente, siempre corresponden a gravedad de Einstein m\'as un campo escalar m\'inimamente acoplado a la curvatura. De esta manera, se puede mostrar que gravedad $f(R)$ es equivalente no s\'olo a una teor\'ia tensor-escalar sino tambi\'en a una teor\'ia de Einstein, acoplada a un fluido ideal, propiedad que es \'util en escenarios inflacionarios \cite{cperalta}.

Adem\'as, el diagrama de Hubble de supernovas tipo Ia, mostr\'o evidencia de que el universo est\'a actualmente en una fase de expasi\'on acelerada,  \cite{supernovapro}. La explicaci\'on m\'as simple, que mejor se ha ajustado a los datos, es el de la constante cosmol\'ogica $\Lambda$. Las observaciones sugieren que existe una especie de fluido desconocido de presi\'on negativa, en la densidad de materia, que domina todo el universo (en un $68\%$ de acuerdo a su medici\'on m\'as reciente \cite{planck}) conocida como energ\'ia oscura a la cual se le asocia la expasi\'on acelerada. Sin embargo, el modelo de constante cosmol\'ogica falla al explicar por qu\'e el valor inferido es peque\~no en comparaci\'on con el valor t\'ipico predicho por la densidad de energ\'ia de vac\'io en el modelo est\'andar de part\'iculas elementales (en 120 \'ordenes de magnitud por debajo lo cual es una cat\'astrofe para la teor\'ia). Una forma de aproximarnos a este problema es asumir que la aceleraci\'on no est\'a aso\-cia\-da directamente al fluido desconocido sino que es una acci\'on neta de la din\'amica del espacio-tiempo. Desde este punto de vista, se intenta modificar las ecuaciones de Einstein desde su geometr\'ia para intentar ajustar los datos reportados sin necesidad de incluir fluidos ex\'oticos en el modelo, es decir, s\'olo con materia est\'andar.

En el presente trabajo se estudia el colapso gravitacional en el marco de la teor\'ia de gravedad mo\-di\-fi\-ca\-da $f(R)$, \cite{borizov}. Este trabajo se divide principalmente en cuatro partes. Inicialmente, se estudian las perturbaciones m\'etricas de primer orden de la teor\'ia usando un modelo de universo de Friedmann-Lema\^{i}tre-Robertson-Walker (FLRW), obteniendo as\'i una ecuaci\'on tipo Poisson, la cual gobierna la din\'amica del perfil de densidad. En segundo lugar, se estudia las ecuaciones de campo en teor\'ias de gravedad modificada $f(R)$. En tercer lugar, se muestra las ecuaciones din\'amicas modificadas. Finalmente, llevaremos la formulaci\'on de perturbaciones de primer orden para obtener la ecuaci\'on tipo Poisson correspondiente en el marco de la teor\'ia de gravedad modificada $f(R)$.
\section{Perturbaciones en Relatividad General}
Tomando un gauge conforme Newtoniano, la m\'etrica de FLRW resultante es 
\begin{equation} \label{metricaperturbada}
ds^2 = a^2(\tau)\left[-(1 +2\phi)d\tau^2 + (1-2\psi)(dx^2 + dy^2 + dz^2), \right]
\end{equation}
donde $\psi$ y $\phi$ son las perturbaciones escalares y
\begin{equation}
\tau \equiv \int^t_0 dt'/a(t'),
\end{equation}
el tiempo conforme. La din\'amica del campo gravitacional est\'a dada por la ecuaci\'on de campo de Einstein  (ver \cite{malik},  \cite{Bardeen} y \cite{hortua})
\begin{equation}
G_{\alpha\beta} = k T_{\alpha\beta}.
\end{equation}
donde $k = 8\pi G$ y $c = 1$ son constantes. Usando una m\'etrica tipo FLRW y tomando perturbaciones de primer orden, las componentes del tensor de Einstein son
\begin{eqnarray}
G_{00} &=& 2\nc\psi - 6\mh\psi' + 3\mh^2,\\
G_{0j} &=& 2\partial_j(\psi' + \mh\F),\label{g0i}\\
G_{ij} &=& (-\mh^2 - 2\mh')\D_{ij} + \big[2\psi'' - \nc(\psi-\F) + \mh(2\phi'-5\psi) \nonumber\\&&- (2\mh' + 4\mh^2)(\phi + \psi) \big]\D_{ij} + \partial_i\partial_j(\psi-\F).
\end{eqnarray}
Adicionalmente, definimos un tensor de Energ\'ia-Momentum para un modelo de polvo
\begin{equation}
T^{\alpha\beta}=(\bar{\rho} +\delta\rho)u^{\alpha}u^{\beta}.
\end{equation}
 La componente $0-0$ de las ecuaciones de campo obtenida es
\begin{eqnarray} \label{estrella}
\D G_{00} &=& 8\pi G \D T_{00}, \nonumber\\
	2\nc\psi - 6\mh\psi'	&=& 8\pi G a^2(\D\R + 2\R\F),\nonumber\\
	\nc\psi &=& 3\mh\psi' + 4\pi G a^2\D\R + 8\pi Ga^2\R\F.
\end{eqnarray}

de la componente $0-j$ se obtiene la condici\'on
\begin{eqnarray} \label{2estrella}
\D G_{0j} &=& 8\pi G \D T_{0j} = 0, \nonumber\\
2\partial_j(\psi' + \mh\F) &=& 0 ,\nonumber\\
 \psi' + \mh\F &=& 0, \nonumber\\
 \psi' &=& -\mh\F.
\end{eqnarray}

Tomando la ecuaci\'on (\ref{2estrella}) y remplazando en la ecuaci\'on (\ref{estrella}) obtenemos
\begin{equation} \label{triangulo}
\nc\psi = -3\mh^2\F + 4\pi G a^2\D\R + 8\pi Ga^2\R\F.
\end{equation}
Por otro lado, la ecuaci\'on de Friedmann en el fondo es
\begin{equation}
\mh^2 = \frac{8\pi G}{3} a^2\bar{\R}.
\end{equation}
Por la ecuaci\'on (\ref{triangulo}) obtenemos finalmente
\begin{eqnarray} \label{poisson}
\nc\psi &=& 4\pi G a^2\D\R.
\end{eqnarray} 
Esta es la ecuaci\'on de campo de Einstein an\'aloga a la ecuaci\'on de Poisson en gravedad Newtoniana. La soluci\'on de la ecuaci\'on (\ref{poisson}) en cordenadas com\'oviles es
\begin{equation}
\psi(x,\tau) = -4\pi G a(\tau)^2 \int  \frac{\delta\rho(x',\tau)}{|x - x'|}d^3x',
\end{equation}
y la ecuaci\'on de movimiento para part\'iculas de materia oscura en tiempo conforme est\'a dada por
\begin{equation}
\frac{d^2\mathbf{x}_c}{d\tau^2} + \mh(\tau)\frac{d\mathbf{x}_c}{d\tau}  = -\nabla_{\mathbf{x}}\psi.
\end{equation}
\section{Ecuaciones de Campo en Teor\'ias de Gravedad Modificada $f(R)$}
La acci\'on general incluye un t\'ermino de frontera $S_{Bond}$, esta se escribe como
\begin{equation}
S_{mod} = \frac{1}{2k}\left(S + S_{Bond}	\right) + S_m,
\end{equation}
con
\begin{equation}
S = \int_{\mathcal{V}} d^4x\sqrt{-g}\left( R + f(R) \right),
\end{equation}
el t\'ermino de frontera descrito como
\begin{equation}
S_{Bond} = 2\oint_{\partial\mathcal{V}}d^3y \epsilon \sqrt{|h|}f_R(R),
\end{equation}
y $S_m$ es la acci\'on de materia. Tomando las variaciones respectivas y evaluando apro\-pia\-da\-mente los t\'erminos en la frontera, obtenemos (ver \cite{guarnizo1}, \cite{guarnizo2})

\begin{equation} \label{ecampo}
G_{\alpha\beta} + f_R(R)R_{\alpha\beta} -\frac12 f(R)g_{\alpha\beta} + g_{\alpha\beta}\square f_R(R) - \nabla_{\alpha}\nabla_{\beta} f_R(R) = kT_{\alpha\beta}. 
\end{equation}
La traza de la ecuaci\'on (\ref{ecampo}) es
\begin{equation}
3\square f_R + f_RR - 2f = kT,
\end{equation}
donde $\square=\nabla_{\alpha}\nabla^{\alpha}$, es el operador D'Alembertiano. 
\section{Ecuaciones Din\'amicas Modificadas}
Las ecuaciones din\'amicas modificadas para la m\'etrica de FLRW perturbada a primer orden y fluido perfecto son
 \begin{itemize}
 \item \textbf{Ecuaci\'on de Friedmann Modificada:} 
 \begin{equation} 
 3H^{2}= \frac{1}{f_R}\left(k\R + \frac12(Rf_R - f) - 3Hf_{RR}\dot{R} \right).
 \end{equation} 
 \item \textbf{Ecuaci\'on de Raychaudhuri Modificada:}
\begin{equation}
2\dot{H}^2 + 3H^{2}= -\frac{1}{f_R}\left(kP + \frac12(f - Rf_R) +f_{RR}\ddot{R} + f_{RRR}\dot{R}^2\right).
\end{equation} 
 \item \textbf{Ecuaci\'on de Continuidad Modificada:}
 \begin{equation}
k\R_m\frac{f_{RR}\dot{R}}{f_R^2} = \dot{\R}_{curv} + 3H\R_{curv}(1 + \omega_{curv}),
\end{equation}
 \end{itemize}
donde se define
\begin{itemize}
  \item $\R_{curv} = \left(\frac{Rf_R- f}{2} - 3Hf_{RR}\dot{R}\right)/f_R$.
  \item $P_{curv} = \left(2H\dot{R}f_{RR} + \frac{f- Rf_R}{2} + \ddot{R}f_{RR} + \dot{R}2f_{RRR} \right)/f_R$.
 \item $\omega_{curv} = \frac{ \ddot{R}f_{RR} + \dot{R}2f_{RRR}-H\dot{R}f_{RR} }{\frac{Rf_R- f}{2}- 3Hf_{RR}t{R}} -1.$
\end{itemize}
aqu\'i el punto indica la derivada respecto del tiempo cosmol\'ogico.

\section{Ecuaciones de Campo Perturbadas con $f(R)$ y Modelo de Polvo}

La ecuaci\'on de campo (\ref{ecampo}) podemos rescribirla as\'i

\begin{eqnarray}\label{campomod2}
(1 +f_R) R_{\alpha\beta} - g_{\alpha\beta}\left( \frac{R + f}{2}
- \square f_R(R) \right) - \nabla_{\alpha}\nabla_{\beta}f_R(R) = kT_{\alpha\beta}.
\end{eqnarray}
Para continuar con nuestro an\'alisis debemos considerar la teor\'ia en el r\'egimen cuasi-est\'atico, en las que tendremos en cuenta las siguientes aproximaciones
\begin{itemize}
\item $\frac{f}{R}<<1$,
\item $f_R << 1$,
\item $f_R R << 1$,
\item $\nabla f_R >> f'_R$.
\end{itemize}
De modo que la ecuaci\'on (\ref{campomod2}) bajo estas aproximaciones queda de la siguiente forma
\begin{equation}
G_{\alpha\beta} + g_{\alpha\beta} \nc f_R - \nabla_{\alpha}\nabla_{\beta}f_R = kT_{\alpha\beta},
\end{equation}
cuya traza tiene la siguiente forma
\begin{equation}\label{trazaaprox}
\nc f_R = \frac13\left[kT + R \right].
\end{equation}

As\'i para la componente $0-0$ tenemos
\begin{equation}
G_{00} + g_{00} \nc f_R - \cancel{\nabla_{0}\nabla_{0}f_R} = kT_{00}.
\end{equation}
Introduciendo las perturbaciones obtenemos
\begin{eqnarray}
\bar{G}_{00} + \delta G_{00} - a^2\left(1+2\phi \right)\left(\frac13\left[-k\left(\bar{\rho} + \delta\rho \right) + \bar{R} + \delta R \right]\right) &=& a^2 k\left(\bar{\rho} + \delta\rho + 2\phi\bar{\rho} \right).
\end{eqnarray}
Sustrayendo el fondo FLRW resulta
\begin{eqnarray}\label{campo00}
2\nc\psi - 6\mh\psi' - a^2\left(\frac13\left[-k\left(\delta\rho + \delta R \right)\right]+ \frac{2\phi}{3}\left[-k\bar{\rho} + \bar{R} \right] \right) &=& a^2 k\left(\delta\rho + 2\phi\bar{\rho} \right). 
\end{eqnarray}
De la componente $0-j$ del tensor de Einstein para el modelo de polvo se obtiene la condici\'on
\begin{equation}
\psi' = -\mh\F.
\end{equation}
Por su parte, la ecuaci\'on de Friedmann modificada a orden cero en el tiempo conforme bajo las condiciones de la aproximaci\'on cuasi-est\'aticaviene expresada como 
\begin{equation}
3\mh^2 = a^2\left(k\bar{\rho} + \frac13\left[-k\bar{\rho} + \bar{R} \right] \right),
\end{equation}
por lo tanto, en la ecuaci\'on (\ref{campo00}) queda escrita como
\begin{equation}
2\nc\psi + \cancel{6\mh^2\phi} - \cancel{2\phi a^2\left(k\bar{\rho} + \frac13\left[-k\bar{\rho} + \bar{R} \right] \right)} = a^2 \left(k\delta\rho  + \frac13\left[-k\delta\rho + \delta R \right]\right).
\end{equation}
De esta manera obtenemos una ecuaci\'on de tipo Poisson modificada en $f(R)$ para el caso cuasi-est\'atico
\begin{equation} \label{poissonmod}
\boxed{\nc\psi = a^2\left(\frac13 k\D\R + \frac16\delta R \right)}.
\end{equation}
Note que en el l\'imite de la relatividad general se obtiene que $\delta R = 8\pi G \delta\rho$. Finalmente la  ecuaci\'on para el potencial gravitacional se reduce la ecuaci\'on no modificada
\begin{eqnarray} 
\nc\psi &=& 4\pi G a^2\D\R.
\end{eqnarray} 
\section{Conclusiones}
En el contexto de la relatividad general, al retirar el fondo y analizar solamente la parte neta perturbada de las ecuaciones de campo de Einstein de primer orden, obtuvimos que sus componentes se presentan en forma de ecuaci\'on tipo Poisson
\begin{eqnarray}
\nc\psi &=& 4\pi G a^2\D\R.
\end{eqnarray} 

Adicionalmente, la ecuaci\'on geod\'esica para materia oscura es
\begin{equation}
\frac{d^2\mathbf{x}_c}{d\tau^2} + \mh(\tau)\frac{d\mathbf{x}_c}{d\tau}  = -\nabla_{\mathbf{x}}\psi
\end{equation}

Por su parte, Las componentes de las ecuaciones de campo en gravedad modificada perturbadas hasta primer orden bajo la aproximaci\'on cuasi-est\'atica producen ecuaciones tipo Poisson que relacionan el potencial de la perturbaci\'on con el elemento de densidad perturbado m\'as un t\'ermino perturbado de la curvatura.

\begin{equation} 
\nc\psi = a^2\left(\frac13 k\D\R + \frac16\delta R \right)
\end{equation}

Al comparar los escenarios de gravedad de Einstein con la de gravedad modificada $f(R)$ bajo la aproximaci\'on cuasi-est\'atica se recuperan las ecuaciones de campo perturbadas de primer orden en Relatividad General.

Esta ecuaci\'on junto a la ecuaci\'on de la traza
\begin{equation}
\nc f_R = \frac13\left[kT + R \right]
\end{equation}

Son las ecuaciones que dominan el colapso gravitacional en este contexto.

\section*{Agradecimientos}

Esta investigaci\'on se llev\'o a cabo en las instalaciones del Observatorio Astron\'omico Nacional de la Universidad Nacional de Colombia. Agradezco al profesor Leonardo Casta\~neda Colorado junto al Grupo de Gravitaci\'on y Cosmolog\'ia por el desarrollo y conclusi\'on de las ideas aqu\'i plasmadas.

\renewcommand{\refname}{Bibliograf\'ia}
\bibliographystyle{harvard}
\bibliography{Cesar}

\end{document}